# Efficient ultrafast photoacoustic transduction on Tantalum thin films

Konstantinos Kaleris[1,2,*], Emmanouil Kaniolakis-Kaloudis[1,2], Evaggelos Kaselouris[1,2], Kyriaki Kosma[1], Emmanouil Gagaoudakis[3], Vassilis Binas[3], Stelios Petrakis[1,2], Vasilis Dimitriou[1,2], Makis Bakarezos[1,2], Michael Tatarakis[1,4], Nektarios A. Papadogiannis[1,2,**]

*kkaleris@hmu.gr

**npapadogiannis@hmu.gr

[1] Institute for Plasma Physics and Lasers (IPPL), Hellenic Mediterranean University, Tria Monastiria, GR-74100 Rethymnon, Greece

[2] Physical Acoustics and Optoacoustics Laboratory, Department of Music Technology and Acoustics, Hellenic Mediterranean University, GR-74100 Rethymnon, Greece

[3] Institute of Electronic Structure and Laser (IESL) of the Foundation for Research and Technology – Hellas (FORTH), Vassilika Vouton, GR-70013 Heraklion, Greece

[4] Department of Electronic Engineering, Hellenic Mediterranean University, GR-73133 Chalepa, Chania, Greece

**Abstract.** Nano-acoustic strain generation in thin metallic films via ultrafast laser excitation is widely used in material science, imaging and medical applications. Recently, it was shown that transition metals, such as Titanium, exhibit enhanced photoacoustic transduction properties compared to noble metals, such as Silver. This work presents experimental results and simulations that demonstrate that among transition metals Tantalum exhibits superior photoacoustic properties. Experiments of nano-acoustic strain generation by femtosecond laser pulses focused on thin Tantalum films deposited on Silicon substrates are presented. The nano-acoustic strains are measured via pump-probe transient reflectivity that captures the Brillouin oscillations produced by photon-phonon interactions. The observed Brillouin oscillations are correlated to the photoacoustic transduction efficiency of the Tantalum thin film and compared to the performance of Titanium thin films, clearly demonstrating the superior photoacoustic transduction efficiency of Tantalum. The findings are supported by computational results on the laser-induced strains and their propagation in these thin metal film/substrate systems using a Two-Temperature Model in combination with thermo-mechanical Finite Element Analysis. Finally, the role of the metal transducer-substrate acoustic impedance matching is discussed and the possibility to generate appropriately modulated acoustic pulse trains inside the crystalline substrate structures for the development of crystalline undulators used for γ-ray generation is presented.

# 1. Introduction

The generation of acoustic pulses on thin metallic films by focused ultrashort laser pulses finds application in a wide range of scientific and technological fields. In medical diagnostics, for example, laser-induced acoustic waves produced on metallic films, are propagated into biological tissues for non-destructive imaging and characterization of single cells [1][2][3], while in material diagnostics, detection of cracks and flaws inside target materials is achieved by capturing reflections of photo-acoustically generated pulses from the material defects [4]. Thin metallic film transducers are also used for the evaluation of the elastic and mechanical properties e.g., thermal conductivity and Young's modulus, of materials of industrial interest [4][5]. For example, in the steel industry, the sensitivity of ultrasonic velocity-to-phase fraction is used for the monitoring of phase transformation of austenite into ferrite. Specifically, it measures the retained austenite fraction in multiphase steels [4]. Other mechanical material properties which can be quantified or monitored via laser ultrasonics are steel annealing, cohesion strength and grain orientation [4][5][6].

Besides their industrial importance thin metallic transducers have found their way into various scientific studies. For example, C. Rossignol et al [8] generated thermoelastic shear acoustic waves using femtosecond laser pulses in submicrometric isotropic aluminum films. V. Bessea et al. [9] studied the generation and exchange of magnons by ultrashort, picosecond scale acoustic pulses propagating through ferromagnetic thin films, acting as transducers. In previous work we have studied the photoacoustic properties of Titanium (Ti) by generating high frequency acoustic strains in Ti/Si (Silicon) samples with ultrafast laser pulses [9], as well as the effect of optical chirp on the efficiency of laser-generated acoustic phonons [10]. Finally, photoacoustic excitation is utilized for the excitation of vibrations in nanorods and nanowires, allowing for their precise thermomechanical characterization [11][12][13].

To implement the described scientific and industrial applications with the required accuracy, various experimental schemes have been developed for the generation and measurement of nano-acoustic strains on thin metal film/substrate systems. One such widely used scheme is the pump-probe transient reflectivity experiment, in which the generated nano-acoustic strain is reflected back and forth between the thin metallic film surfaces [14][15][16]. By measuring this reflection echo and/or the Brillouin oscillation frequency, a detailed acousto-mechanical characterization of the material system is achieved. More specifically, interaction of ultrashort laser pulses with metallic films leads to the deposition of optical energy in a region close to the film's surface causing a rapid heating. This is followed by a localized thermal expansion and consequent elastic contraction of the metal lattice that results in the generation of an acoustic strain in the form of a N-pulse. Typically, such a highly localized acoustic strain has a duration of the order of several tens of picoseconds and a longitudinal spatial spread of several tens of nanometers. The strain propagates away from the metal/substrate interface and penetrates into the substrate, partially reflected back towards the metal's surface. Assuming normal incidence at the boundary, the amount of transmitted and reflected acoustic energy is calculated in terms of the acoustic impedances of the metallic film and the material by the Fresnel equations:

$$R = \frac{(Z_2 - Z_1)}{(Z_2 + Z_1)}, \qquad T = \frac{2Z_2}{(Z_2 + Z_1)} \tag{1}$$

where $R, T, Z_1, Z_2$ are the reflection and transmission coefficients and the acoustic impedances of the metallic film and substrate material of interest, respectively.

The quality of a metallic film as a photoacoustic transducer is determined by its dynamic behaviour to thermal excitation and depends on a variety of optical and thermo-mechanical properties of the metal as well as the characteristics of the laser radiation. The efficiency of the electronic excitation depends on the metal's band structure and the laser characteristics, which determine the probability of an electron transition to a higher energy state. Excitation is followed by electron-electron and electron-phonon interactions that collectively lead to the thermalization of the metallic film. As such, the quality of the photoacoustic transducer strongly depends on the electron-phonon coupling constant $G$, which describes the strength of the electron-phonon interaction and, consequently, the amount of energy transferred from the excited electrons to the lattice. There are many other parameters that strongly affect the transduction efficiency of the metal, such as the thermal expansion coefficient $a_V$, which describes the volume change per unit temperature change, the Young's modulus $E$, which governs the relation between stress and strain in the lattice, the specific lattice heat capacity, which describes the energy per unit mass needed for a unit change in the temperature, the yield strength, which determines the maximum allowed stress within the elastic regime of the metal, and the melting point. Considering also that most parameters are temperature dependent, it becomes apparent that estimation of the transduction efficiency of metallic films is demanding.

In this work, we evaluate Tantalum (Ta) thin films regarding their performance as photoacoustic transducers, using ultrashort laser pulse excitation. Transition metals such as, Ti have been shown in previous work to exhibit superior photoacoustic transduction properties compared to noble metals such as Silver (Ag) [7]. Here, the excellent photoacoustic transduction properties of Ta, which is also a transition metal, and its suitability for the efficient production of high-amplitude ultrashort nano-acoustic strains is demonstrated by comparison to the performance of Ti. Moreover, the effect of the acoustic impedance matching between metal and substrate on the strains propagating in the substrate is discussed. It is shown that appropriate tuning of the material impedances and the thickness of the metallic film allows for the formation of acoustic pulse trains with controlled spatiotemporal characteristics inside the substrate. A potential application of such pulse trains propagating in crystalline materials for the development of crystalline undulators (CUs) is presented and analysed. In this application the strain modulation of the crystal lattice enables the undulation of an ultra-relativistic positron beam for the generation of brilliant, narrow bandwidth γ-ray radiation that could be especially useful for medical applications. Theoretical work on such crystalline light sources (CLS) has been carried out by Solov'yov and Korol [17][18][19][21], while preliminary experimental evaluations have been recently carried out by various groups [22][23][24][25].

This paper is structured as follows: in Section 2, the experimental and computational methodology for the evaluation of the photoacoustic properties of Ta and Ti thin films is presented. In Section 3 the comparative experimental and computational results for the photoacoustic transduction properties of Ta and Ti, and the propagation of the nano-acoustic strains in Si, are presented and discussed. In Section 4, an application scheme of the ultrafast generation of acoustic pulse trains inside crystalline materials via metallic photoacoustic transducers for the development of CUs is proposed and discussed. Finally, Section 5 presents the summary and conclusions of this work.

## 2. Methodology

This work follows a combined experimental and computational methodology for quality evaluation of Ta thin films as photoacoustic transducers. For the experiments, the pump-probe transient reflectivity technique is used, while for the simulations, the Finite Element Analysis (FEA) method is adopted combined with a Two-Temperature Model (TTM) accounting for the initial non-thermal electron excitation by the ultrashort laser pulses. The methods are analyzed in detail in the following subsections.

### 2.1. Experimental measurements

Pump-probe transient reflectivity is a well-established technique for thin film metrology that allows for the generation and high-precision monitoring of acoustic strains as they propagate inside layered materials [14][15][16]. In this method, probing relies on the Brillouin scattering of photons due to interaction with the propagating phonon at different depths inside the material. More specifically, after excitation by the pump beam, a delayed probe beam measures the changes in the sample's overall reflectivity induced by the pump [14][15]. The probe beam energy is sufficiently low so that it does not induce significant thermoelastic excitation of the metal. As shown in Fig. 1.a, part of the probe beam is initially reflected while the rest is transmitted through the metal where it is progressively absorbed at a rate governed by the metal's absorption coefficient. For a few tens of nanometers thick metallic films deposited on a substrate (material of interest), part of the transmitted probe beam enters the substrate and interacts with the induced phonon inside the substrate at a depth that depends on the relative delay between pump and probe. The probe photons are scattered by the phonon in a process called Brillouin scattering [16], which involves absorption of a photon by the lattice and consequent re-emission at a shifted frequency, where the energy difference is gained by the phonon. The back-scattered probe photons finally exit the layered material and interfere with the initially reflected part of the probe beam. The composite probe beam is finally captured by a photodiode generating the transient reflectivity signal $S_{\Delta R/R}$. This process is repeated for a range of pump-probe time delays $\Delta t$ starting shortly before $\Delta t = 0$ and extending up to a hundred picoseconds or more.

In the $S_{\Delta R/R}$ signal (see also Fig. 2), two main features are imprinted: a prominent peak around $\Delta t = 0$ corresponding to the electronic excitation of the metallic film and a slowly decaying tail modulated by the so-called Brillouin oscillations [16]. The Brillouin oscillations are the result of constructive or destructive interference between the interfering parts of the probe beam depending on the optical path traversed by the back-scattered beam. It can be easily shown that the oscillation frequency $f_{Br}$ is given by [7]:

$$f_{Br} = \frac{2n_s u_s}{\lambda} \tag{2}$$

where $n_s, u_s$ and $\lambda$ are the refractive index and the longitudinal speed of sound in the substrate respectively, and $\lambda$ is the laser wavelength. Significantly, assuming the same substrate, the modulation depth of the Brillouin oscillations is directly related to the amplitude of the induced phonon. This particular aspect of the transient reflectivity signal is used here for the evaluation of Ta photoacoustic transduction properties.

A diagram of the used experimental setup is shown in Fig.1. The output of a femtosecond Ti:Sapphire laser system from FEMTOLASERS Produktions GmbH (Femtopower compact PRO) with a pulse duration of 35 fs and of wavelength of 800 nm operating at a repetition rate of 1 kHz is split into two parts using a beam splitter, where 90% of the pulse energy goes to the pump line and the remaining 10% to the probe line. The pump beam is focused onto the sample using a thin converging lens with a focal length of 25 cm exciting the metal transducer at a modulation frequency of 270 Hz, as imposed by a mechanical chopper.

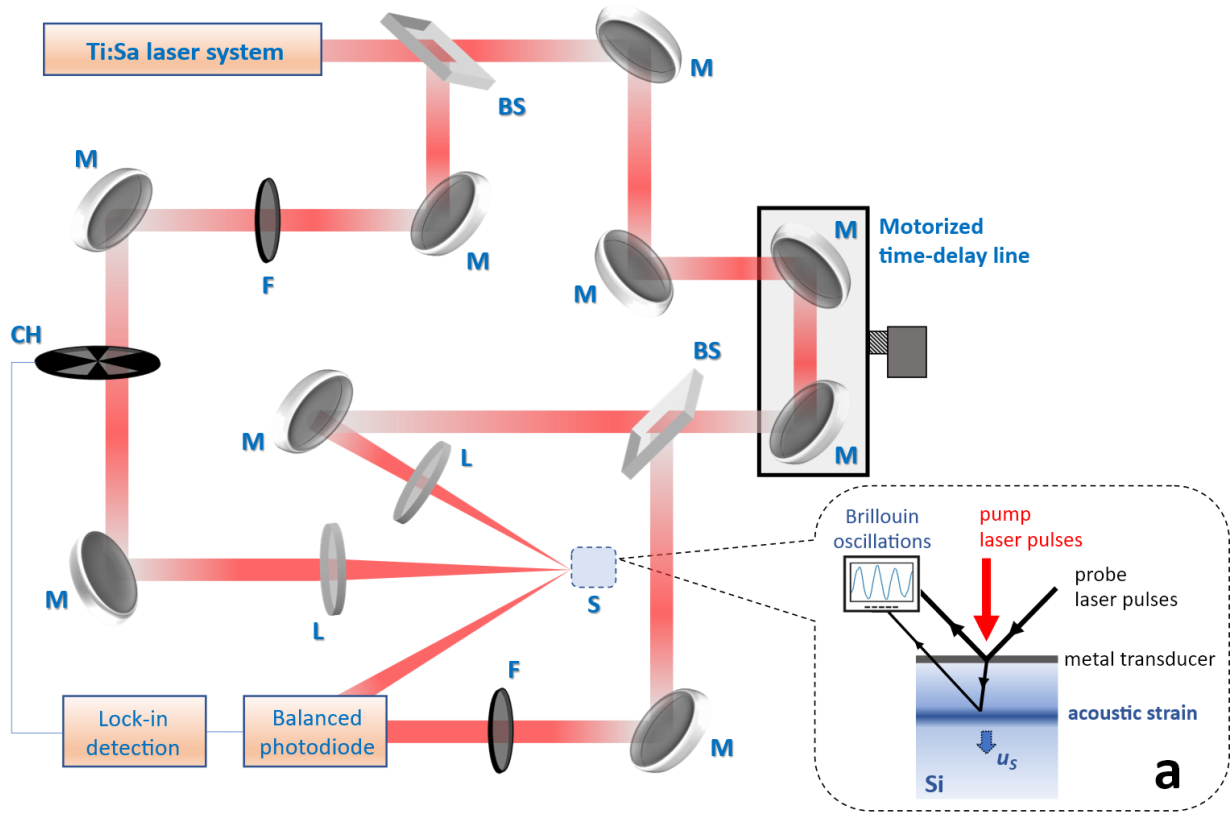

Fig.1: Experimental setup used for the evaluation of the photoacoustic properties of Ta thin films, based on the pump-probe transient reflectivity method: L: thin lens, M: mirror, BS: beam splitter, CH: chopper, S: sample, F: neutral density filter. The inset (a) is a schematic of the induced strain inside the sample and the corresponding reflections of the probe beam.

The energy of the pump pulse is ~30 μJ, which corresponds to a fluence of ~50 mJ/cm$^2$. The energy and fluence were selected so that the obtained transient reflectivity signals from both samples were maximized, without having any visibly or experimentally perceptible detrimental effects on the metal coatings. The progressive time delay between the pump and probe beams is introduced by passing the probe beam through a retroreflecting translation stage at a time step of 30 fs. Before reaching the sample, the probe beam passes through a 50:50 beam splitter with one part directed to the reference input of a balanced photodiode and the other part focused onto the sample using a lens with a focal length of 15 cm. The reflected probe beam enters the second input of the balanced photodiode. The output signal of the photodiode is directed into a dual channel lock-in amplifier which captures the reflectivity changes at the pump modulation frequency. A camera is used to monitor the sample surface, dimensions and position of the pump and probe beams. The components of the pump-probe set-up including the translation stage, camera, and lock-in amplifier, as well as the parameters of the experiment are controlled by specially designed software.

Finally, both the Ta/Si and Ti/Si samples have metallic films of 25 nm thickness while the Si <100> substrates are 500 μm thick. The metal deposition on the substrates was done via dc sputtering, using an Alcatel sputtering system and the respective metallic targets. Prior to the deposition, the chamber was evacuated down to $10^{-6}$ mbar. Argon (Ar) gas was used for plasma discharge. The pressure during deposition was kept constant at $4 \times 10^{-2}$ mbar, with a sputtering current of 0.45 A. The thickness was determined by a Veeco Dektak 150 profilometer.

## 2.2. Computational model

To reliably evaluate and compare the photoacoustic behavior of Ta and Ti, lattice temperature gradients and time-dependent strains need to be calculated at various depths inside the layered material. For this purpose, a numerical model is developed here [26][27][28] in the LS-DYNA software [29] based on FEA. Specifically, a TTM is adopted that describes the energy deposition to the multilayered target by ultrafast laser pulses [30]. The sample is modeled using two domains, one for the electrons and one for the lattice, which are identical in terms of geometry and discretization. For the electron domain, only thermal analysis is carried out, while for the lattice domain, thermo-mechanical analysis is performed. The governing equations for are solved simultaneously for both domains, while the heat transfer between the domains is governed by the electron-phonon coupling term via an in-house developed code. The following set of equations is employed to study the electron $T_e$ and phonon temperature $T_p$ based on the TTM:

$$C_e(T_e)\frac{\partial T_e(r,z,t)}{\partial t} = \nabla[k_e(T_e)\nabla T_e(r,z,t)] - G(T_e,T_p)(T_e-T_p) + \frac{\partial U_{ee}}{\partial t} \qquad (3)$$

$$C_p(T_p)\frac{\partial T_p(r,z,t)}{\partial t} = G(T_e,T_p)(T_e-T_p) + \frac{\partial U_{ep}}{\partial t} \text{ (for metallic film)} \qquad (4)$$

$$C_p(T_p)\frac{\partial T_p(r,z,t)}{\partial t} = \nabla[k_p(T_p)\nabla T_p(r,z,t)] + TR\frac{\partial U_{ep}}{\partial t} \text{ (for substrate)} \qquad (5)$$

$$\frac{\partial}{\partial t}\begin{Bmatrix}U_{ee}\\U_{ep}\end{Bmatrix} = \frac{2A\sqrt{ln2}F}{\sqrt{\pi}(hv)^2 t_p}e^{-2(\frac{r}{r_b})^2} \times \int_0^t \left[\frac{1}{1-e^{-\frac{d_m}{a_b^{-1}}}}\frac{1}{a_b^{-1}}e^{-4ln2(\frac{t'-t_0}{t_p})^2}e^{-\int_0^z \frac{1}{a_b^{-1}}dz'}\begin{Bmatrix}H_{ee(t-t')}\\H_{ep(t-t')}\end{Bmatrix}\right]dt' \qquad (6)$$

where $k_e$ is the electron thermal conductivity, $C_e$ and $C_p$ are the heat capacity of electrons and lattice, respectively, $G$ is the electron-phonon coupling factor, $A$ is the absorbance of the laser energy, $F$ is the fluence of the laser beam, *hv* is the one-photon energy, $t_p$ is the laser pulse duration, $r_b$ is the laser beam radius, $d_m$ is the thickness of the metal film, $α_b$ is the absorption coefficient, while it holds $t_0$=-3$t_p$. Moreover, $H_{ee}$ and $H_{ep}$ are functions that contain parameters related to the transient creation of the non-thermal electron distribution [31]. Eq. (3) is applied on the electron domain while Eq. (4) and (5) apply on the lattice domain. Eq. (3) holds for the metallic film while Eq. (4) holds for the Si substrate. The terms $\partial U_{ee}/\partial t$ and $\partial U_{ep}/\partial t$ correspond to the energy densities per unit time transferred by the laser source to non-thermal and thermal electrons, which in turn transfer energy to the lattice. These terms also account for the dynamic behavior of the absorption coefficient [31][32], while $TR$ of Eq. (5) corresponds to the transmitted percentage of the energy density per unit time to the lattice substrate. Furthermore, the conservative equations of mass, momentum and energy are also solved for the lattice domain [27].

The developed FEA models are axisymmetric pseudo-3D with dimensions of the solid target $(X, Y, Z) =$ 180 μm × 1 nm × 1.25 μm. The solid element has a thickness of 1 nm in the Y direction. The metallic films are modeled in their entire depth of 25 nm, while the substrate is modeled only for the first 1.25 μm. The solution domain is discretized by $10^6$ elements, providing mesh-independent simulation results. Non-reflecting boundary conditions are set on the outer areas except for the irradiated area. The full width at half maximum (FWHM) radius of the laser beam on the sample surface $r_b$ is 150 μm, according to the experiments.

Regarding the interface between the Ti or Ta transducer and the Si substrate, a single mesh describes the computational volume of both materials. Since the sample fabrication method leads to a good adhesion between the coated metallic films and the Si substrate, in the model, the metallic film shares the same nodes with the Si substrate at the interface. The same modelling methodology has been applied in relevant previous works [27]. Thus, the heat is transported between the two materials neglecting thermomechanical discontinuities at the interface, so that the temperature and heat flux are modeled as continuous functions. The heat transport along the metal-substrate interface is modeled through the term $TR \frac{\partial U_{ep}}{\partial t}$, as shown in Eq. (5).

Regarding the thermal material properties of Ti, the data of the temperature dependent electron heat capacity $C_e$ and electron-lattice coupling factor *G* found in [32] were fitted using a first order polynomial based on the work in [7]. The temperature-dependent electron heat capacity is given by:

$$C_e(T_e, T_p)^{Ti} = C_{e0}^{Ti} \frac{T_e}{\frac{A_e}{B_p}(T_e)^2 + T_p} \tag{7}$$

where, $C_{e0}^{Ti}$ is the electron heat capacity in room temperature, and the coefficients $A_e$ and $B_p$ are constants. The equation of the temperature dependent electron-lattice coupling factor is:

$$G(T_e, T_p)^{Ti} = G_0^{Ti}\left[\frac{A_e}{B_p}(T_e + T_p) + 1\right] \quad (8)$$

where, $G_0^{Ti}$ is the electron–phonon coupling strength in room temperature, and the coefficients $A_e$ and $B_p$ are constants. The electron thermal conductivity $K_e$ for Ti is calculated from the heat capacity, the electron-electron collision frequency, the electron-phonon collision frequency and the Fermi velocity [33][34]. For the $K_e$ it holds

$$K_e(T_e, T_p)^{Ti} = \frac{2E_F}{m_e}\frac{C_e^{Ti}}{3(A_e T_e^2 + B_p T_p)} \quad (9)$$

where $E_F$ is the Fermi energy (8.84 eV) and $m_e$ the electron mass. For Ta the values of the electron thermal parameters $C_e^{Ta}$, $K_e^{Ta}$ and $G^{Ta}$ can be found in [35], where $C_e^{Ta} = C_{e0}^{Ta}T_e$, $K_e^{Ta} = K_0\frac{T_e}{T_p}$. The relevant expressions for the electron thermal parameters of Si can be found in [34], where $C_e^{Si} = \frac{3}{2}k_B n_e^{Si}$, $K_e^{Si} = 2k_B^2 \mu_e^{Si} n_e^{Si} T_e e^{-1}$, where $k_B$, $n_e$, $\mu_e$ and $e$ are the Boltzmann constant, the electron density, the electron mobility and the electron charge, respectively.

The mechanical properties such as density $\rho$, Young's modulus $E$, Poisson ratio $v$ and the yield strength $Ys$ of Ti, Ta and monocrystalline Si <100> can be found in [36] and [37]. Anisotropic mechanical properties are considered for Si, and the values of the material elastic constants ($C_{11}$, $C_{12}$, $C_{44}$) used in the simulations are taken from [38]. Temperature dependent values for all three materials of heat conductivity $K_L$ [39], heat capacity $C_L$ [40] and thermal expansion coefficient $\alpha$ [41] are also considered for the lattice domain. Table 1 presents thermal and mechanical parameters used in the simulations.

**Table 1** Parameters used in the simulations.

| Parameter | Value | Parameter | Value | Parameter | Value |
|---|---|---|---|---|---|
| $A_e$ [$K^{-2}s^{-1}$] | 6.16 × 10$^4$ | $C_{11}^{Si}$ [$Pa$] | 166.2 × 10$^9$ | $Ys^{Si}$[$Pa$] | 0.35 × 10$^9$ |
| $B_p$ [$K^{-1}s^{-1}$] | 1.21 × 10$^{13}$ | $C_{12}^{Si}$ [$Pa$] | 66.4 × 10$^9$ | $\alpha^{Ti}$ [$K^{-1}$] | 8.5 × 10$^{-6}$ |
| $C_{e0}^{Ti}$ [$J\,K^{-2}m^{-3}$] | 319.8 | $C_{44}^{Si}$ [$Pa$] | 79.8 × 10$^9$ | $\alpha^{Ta}$ [$K^{-1}$] | 6.4 × 10$^{-6}$ |
| $C_{e0}^{Ta}$ [$J\,K^{-2}m^{-3}$] | 77.4 | $\rho^{Ti}$ [$kg\,m^{-3}$] | 4500 | $\alpha^{Si}$ [$K^{-1}$] | 2.6 × 10$^{-6}$ |
| $K_0$ [$W\,m^{-1}K^{-1}$] | 57.6 | $\rho^{Ta}$ [$kg\,m^{-3}$] | 16650 | $K_L^{Ti}$ [$W\,m^{-1}K^{-1}$] | 21.9 [at 300K] |
| $G_0^{Ti}$ [$W\,m^{-3}K^{-1}$] | 1.4 × 10$^{18}$ | $\rho^{Si}$ [$kg\,m^{-3}$] | 2324 | $K_L^{Ta}$ [$W\,m^{-1}K^{-1}$] | 57.5 [at 300K] |
| $G_0^{Ta}$ [$W\,m^{-3}K^{-1}$] | 3.1 × 10$^{17}$ | $v^{Ti}$ | 0.323 | $K_L^{Si}$ [$W\,m^{-1}K^{-1}$] | 157 [at 300K] |
| $n_e^{Si}$ [$m^{-3}$] | 1.5 × 10$^{16}$ | $v^{Ta}$ | 0.35 | $C_L^{Ti}$ [$J\,kg^{-1}K^{-1}$] | 522 [at 300K] |
| $\mu_e^{Si}$[$m^2\,V^{-1}s^{-1}$] | 0.15 | $v^{Si}$ | 0.279 | $C_L^{Ta}$ [$J\,kg^{-1}K^{-1}$] | 142 [at 300K] |
| $E^{Ti}$ [$Pa$] | 116 × 10$^9$ | $Ys^{Ti}$[$Pa$] | 0.38 × 10$^9$ | $C_L^{Si}$ [$J\,kg^{-1}K^{-1}$] | 710 [at 300K] |
| $E^{Ta}$ [$Pa$] | 186 × 10$^9$ | $Ys^{Ta}$[$Pa$] | 0.705 × 10$^9$ | | |

Based on the work in [7] the Transfer Matrix Method is used to calculate the total reflectance, absorbance, and transmittance of multi-layered materials, considering the optical admittance of every different medium, namely, the ratio of the amplitudes of the tangential components of magnetic and electric field vectors and the thickness of the metallic layer. Three media are considered: the air, the thin metallic layer and the Si substrate, and thus two interfaces (air/metal and metal/Si) exist. the complex refractive index and absorption coefficient of Ti, Ta and Si at 800 nm can be found in [42] and [43], respectively.

## 3. Results

Fig. 2 presents typical comparative transient reflectivity signals from Ta and Ti thin films coated on Si <100> substrates. At early times, prominent peaks of the reflectivity change due to electronic excitation can be seen, where the peak is significantly stronger in the Ti/Si sample, also decaying faster. Electronic excitation and the resulting transient reflectivity signal have complex dependencies from various parameters including the metal's band structure in relation to the exciting laser wavelength and the electron and lattice temperatures. As a result, a qualitative description of the observations in the electronic excitation part of the transient reflectivity signal is impossible. It should be noted that, a similar behavior is observed in [7], where the electronic excitation signal of Ti is significantly lower than that of

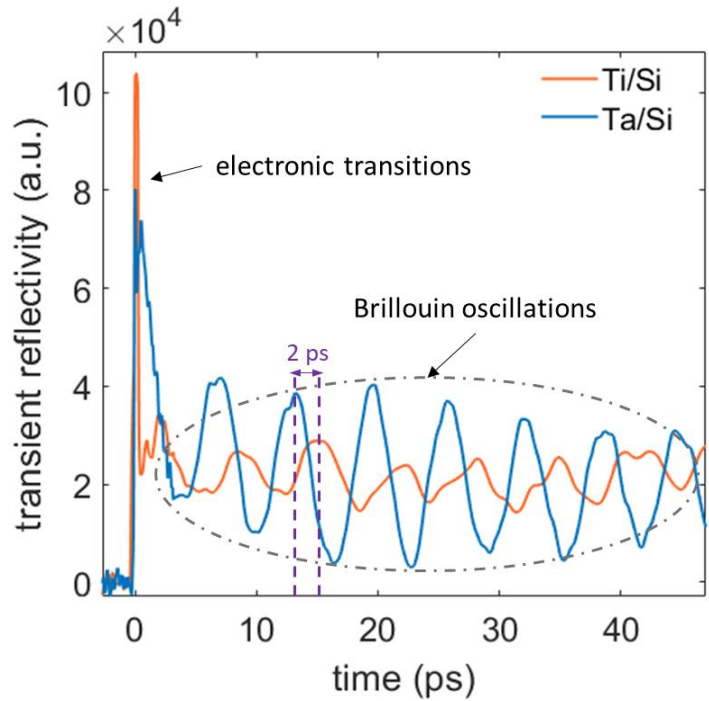

Fig. 2: Typical measured transient reflectivity signals for the 25 nm Ta and Ti films on Si substrates showing the electron excitation and Brillouin oscillations characteristic features.

Ag. However, since this work focuses on the transduction efficiency of Ta which, as shown below, reflects on the latter part of the transient reflectivity signal, the early part of the metal's response will not be further investigated.

At later times, the decaying Brillouin oscillations can be seen, which occur due to the alternating constructive and destructive interferences between the initially reflected and back-scattered probe beam components. From the figure it becomes clear that the modulation depth of Ta/Si is more than two times larger compared to that of Ti/Si. It should be noted that Ta and Ti have almost equal absorption coefficients at 800 nm ($a^{Ta} = 1.15 \times 10^6$ cm$^{-1}$, $a^{Ti} = 1.07 \times 10^6$ cm$^{-1}$) so that there are no significant differences in the transient reflectivity signals due to absorption of the probe beam inside the metals. Also, while the frequency of the Brillouin oscillations is equal for the two samples, as expected due to the common substrate material, a time-difference of about 2 ps in the oscillations of the Ta/Si sample can be observed. This originates from the different longitudinal speeds of sound in Ti ($v^{Ti} = 6070$ m/s) and Ta ($v^{Ta} = 4100$ m/s). Assuming that the photoacoustic transduction takes place on the surface of the metallic films, the travel times of the acoustic pulses inside the 25 nm films are $t^{Ti} = 4.1$ ps and $t^{Ta} = 6.1$ ps, respectively. Hence, the delayed penetration of the acoustic pulse in the substrate for the Ta/Si sample results to a respective delay in the measured Brillouin oscillations.

Regarding the modulation depth in the Brillouin oscillations, it can be argued that the reason behind the larger modulation depth on the Ta sample is the result of enhanced photoacoustic transduction compared to Ti. A qualitative analysis of the observed signals can be outlined based on the parameters described in the Introduction section. Particularly, although Ti has a larger thermal expansion coefficient at room temperature ($a^{Ta} = 6.4\ 10^{-6}$/K, $a^{Ti} = 8.5\ 10^{-6}$/K), and hence a larger expansion per unit temperature change, it also has a significantly larger specific heat capacity (~4 times), meaning that more heat is required for its temperature to change. Importantly, Ta has a larger Young's modulus ($E^{Ta} = 186$ GPa, $E^{Ti} = 116$ GPa) so that it develops higher stresses compared to Ti for the same strain. Hence, it is expected that an equal amount of energy delivered to the Ti and Ta lattices will lead to significantly higher pressures on the Si substrate for the Ta/Si sample. Moreover, Ta has ~2 times higher yield strength and melting point, which allows for the bearing of higher strains. Experimentally, these were found to play an important role in relatively high laser fluences, such as the ones set to maximize the transient reflectivity signals, since Ti is driven close to the plastic region while Ta remains well inside the elastic region. On the contrary, evaluations of the electron-phonon coupling constants $G$ from previous works have shown that $G^{Ti}$ is ~3 times larger than $G^{Ta}$ at room temperature ($G^{Ta} = 3.1 \times 10^{17}$ W/m³/K, $G^{Ti} = 1.4 \times 10^{18}$ W/m³/K) [44][45].

A clear picture of the optoacoustic transduction process and strain propagation inside the samples is provided by the simulations. Fig. 3 shows the model predicted dynamic electron and lattice temperatures under the influence of the pump pulse evaluated at the center of the Ta/Si and Ti/Si samples surface. The electron cloud reaches a higher temperature for Ta sample, with $T_{e,\max}^{Ta} \approx 5800$ K and $T_{e,\max}^{Ti} \approx 4500$ K indicating favorable electronic transitions at 800 nm compared to Ti. Moreover, Ta exhibits slower lattice thermalization, as expected due to the lower electron-phonon coupling factor $G$ of Ta, which results to slower energy transfer from the thermal electrons to the lattice. Finally, the maximum lattice temperatures in the two metals are very close, for Ta being greater by ~100K; they are reached at different times, particularly at ~1.5 ps in Ti and at ~5 ps in Ta, a difference related again to the different $G$ factors.

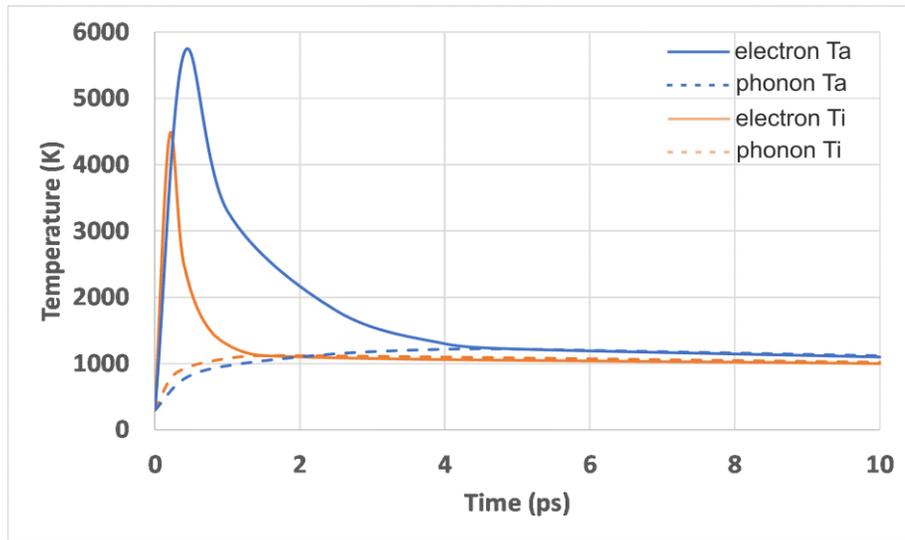

Fig. 3: Calculated temporal evolution of electron (solid lines) and phonon (dashed lines) temperatures at the surface of a 25nm Ta thin film, and a 25nm Ti thin film on Si substrates.

Fig. 4 shows the induced strains inside the two samples at different times, from which several conclusions can be drawn. First, the strain is stronger in Ta as observed in the response of the early times of Figs. 4a) and b). At later times (Figs. 4c) and d)) it can be seen that, for the case of Ta/Si, a significant amount of acoustic energy is reflected at the interface with the substrate.

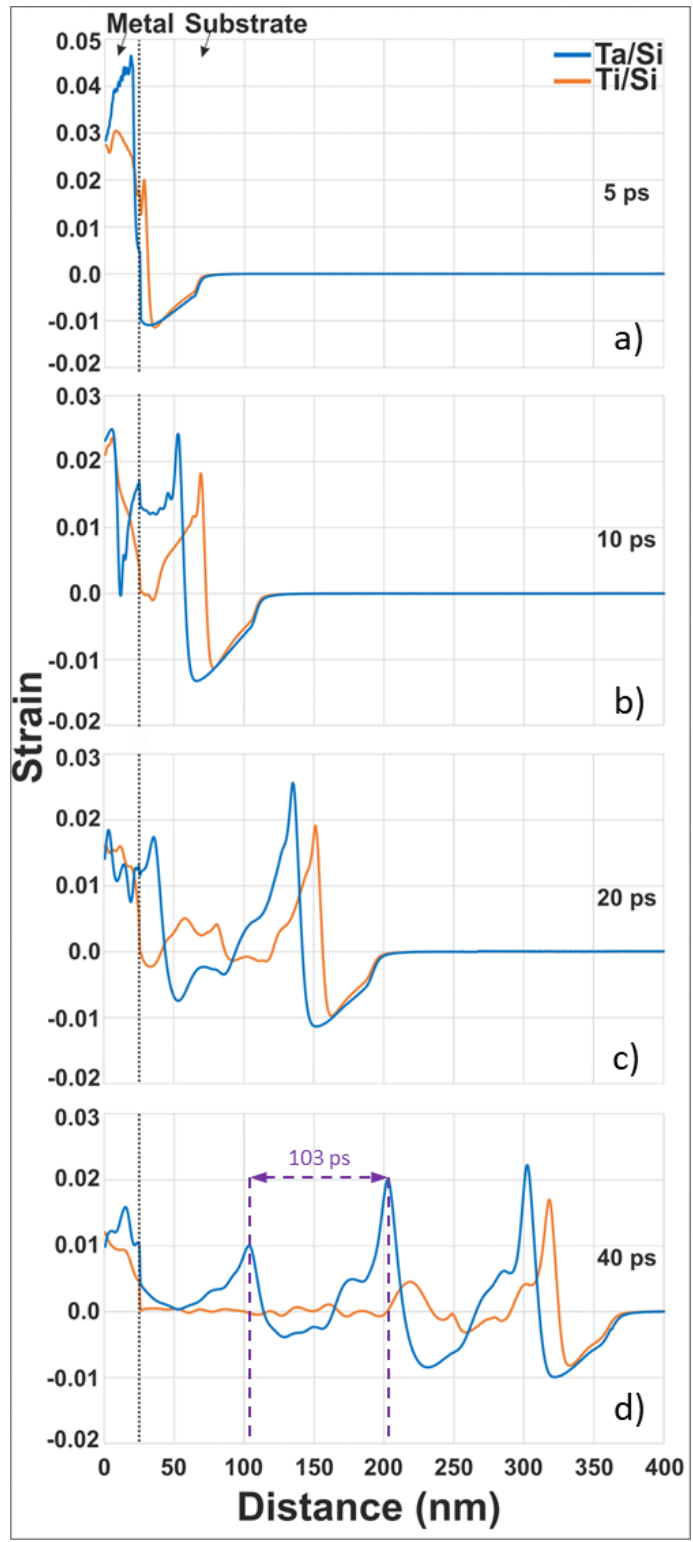

Fig. 4: Calculated strain distribution in 25 nm metal/Si samples as a function of depth, at different time instants. Vertical dashed line corresponds to the position of the metal/Si interface.

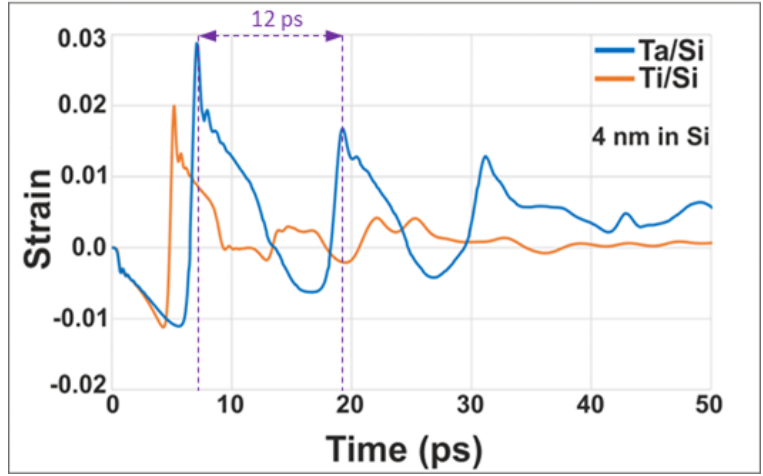

Fig. 5: Calculated strain distribution in 25 nm metal/Si systems as a function of time, for a depth of 4 nm in the Si substrate.

This energy is re-reflected on the film's surface and redirected towards the substrate, leading to the formation of a strain pulse train inside Si. The same conclusion can be drawn from Fig. 5, which shows the time evolution of the strain penetrating the substrate at 4 nm depth from the interface with the metal, as calculated by the model. Looking at the strain in Ta, a first pulse arrives at around 7 ps, while a second and a third pulse arrive at about 19.5 ps and 32 ps respectively, indicating a periodicity of the pulse strains of approximately 12 ps. Considering the longitudinal speed of sound in Ta $v^{Ta} = 4100$ m/s, the duration of one roundtrip of the acoustic pulse inside the 25 nm film is ~12.2 ps. By conversion to a spatial distance inside the Si substrate ($v^{Si} = 8430$ m/s), we get a pulse-to-pulse strain distance of ~103 nm for consecutive strain pulses of the train. These values are in accordance with Fig. 4d) and Fig. 5 of the spatial and temporal difference between the acoustic pulses in the pulse train, respectively.

Moreover, both Figs. 4 and 5 show a progressive decrease in the amplitude of consecutive strains, which is mostly apparent for the Ta/Si sample. For the Ti/Si sample, the primary pulse is dominant while the secondary pulses are significantly suppressed. To interpret these observations, the impedance matching between the thin metallic films and the substrate needs to be investigated for the Ta/Si and Ti/Si samples. Particularly, Ti and Si exhibit a very good impedance matching, since $Z^{Si} \cong 20$ MRayls while $Z^{Ti} \cong 27$ MRayls, which results, according to Eq. (1), to the transmission of more than 85% of the phonon energy into the substrate, with only 15% reflected back into the metal. On the contrary, in the Ta/Si sample ($Z^{Ta} \cong 55$ MRayls) more than 43% of the phonon energy induced in Ta is reflected back. The reflected wave propagates towards the metal's surface, on which it is almost entirely re-reflected towards the substrate, since the very low acoustic impedance of the air ($Z^{air} = 4 \times 10^{-4}$ MRayls) renders the transmission negligible. Thus, in case of sub-optimal impedance matching, the metallic film acts as an acoustic cavity where oscillating acoustic energy is trapped, partially leaving the cavity and propagating into the substrate once in each oscillation. This periodic process leads to the formation of a strain pulse train in the substrate with the time distance between consecutive pulses determined by the speed of sound and the thickness of the metallic film. The energy difference between consecutive pulses is determined by the transmissivity coefficient T at the metal / substrate interface. In the following section, a potential application of such strain pulse trains for the acoustic modulation of electron / positron CUs is presented.

## 4. Proposed application scheme: laser-driven dynamic acoustic crystalline undulators

Pulsed acoustic excitation of crystalline materials with precise control of the spatiotemporal periodicity of the induced acoustic pulses has potential application, among others, in the development of Crystalline Undulators (CUs) for the generation of brilliant, narrow bandwidth γ-rays via undulation of electron ($e^-$) or positron ($e^+$) ultra-relativistic beams. Undulation, namely the forced periodic motion, of charged particle beams propagating at almost the speed of light leads to the emission of coherent photons up to few MeV [19][20] and is currently achieved by the use of strong periodic magnetic fields. Such magnetic fields are commonly produced by large and costly magnets in setups known as Free Electron Lasers (FELs). FELs are housed inside large laboratories, as for example the X-ray Free Electron Laser (European XFEL) of the Deutches Elektronen-Synchron (DESY) [46], which require considerable resources for their operation.

CUs are novel devices that exploit the x extremely high inherent electric fields inside crystal lattices to induce undulation of ultra-relativistic $e^-/e^+$ beams. Such electric fields can reach strengths of up to $10^{10}$ V/cm, or, equivalently, 3000 T, which is massive compared to the maximum of 10 T magnetic fields of typical undulators [19][20]. In CUs, undulation can be achieved by the sinusoidal modulation of the lattice geometry, which can be either static, via appropriate structural deformation, e.g. using periodic mechanical stress, grooving, epitaxially grown superlattices etc. [19][20][47][48] or dynamic, via the generation of acoustic waves inside the crystal [49]. When the $e^-/e^+$ enter the crystal lattice, they get trapped – under conditions described in [17] – within the channels formed by the lattice planes in a process known as $e^-/e^+$ channeling [19][20]. The imposed sinusoidal modulation of the lattice planes forces the charged particles to follow sinusoidal-like trajectories, leading to the emission of narrowband γ-ray photons with energies that can reach several tens of MeV [20]. The propagation length along which the charged particles are expected to be trapped within the lattice channels, known as the channeling length, is a critical parameter that depends on the beam energy and the particle type.

For acoustic wave CUs, periodic pressure fields can be generated inside crystals by means of a transducer, but also by laser-induced periodic acoustic pulse trains, like the ones presented in this work. In this particular scheme depicted in Fig. 6, the beam enters the crystal, e.g. Silicon <100>, diagonally, and propagates through the channels shaped in the <110> crystallographic direction. It should be noted that other crystalline materials, mainly monocrystals that exhibit homogeneous electric fields in the lattice, e.g. Germanium (Ge) and diamond, are also suitable for the development of CUs. Depending on the spatial periodicity of the pulses and the displacement amplitude of the lattice planes, different undulation regimes are defined, as described in [21]. For the Ta/Si sample studied here, the strain pulse-to-pulse distance is 103 nm and the maximum lattice plane displacement is of the order of 0.4 - 0.5 nm, corresponding to the so-called Short Amplitude Short Period (SASP) regime [21][50]. Investigation of the SASP undulation regime for the generation of γ-rays has been previously carried out e.g. by [51], where it was shown that this regime is highly efficient for lower $e^-/e^+$ beam energies, especially due to the large number of periods fitting within the channeling length of the beam.

Importantly, the pulse periodicity and the displacement of the lattice planes can be controlled by tuning the thickness of the metal film, the crystalline material and the characteristics of the laser excitation. For example, for the same Ta coating, the strain pulse-to-pulse distance inside a Ge <100> crystal would be 66 nm while in diamond <100> 146 nm. Accordingly, a 100 nm thick Ta film would result to 400 nm strain pulse-to-pulse distance inside Si, 264 nm in Ge and 584 nm in diamond. Moreover, the strain pulse-to-pulse amplitude difference depends on the impedance matching between the metal transducer and the crystal (Eq. (1)) while the duration of the acoustic strain can be controlled by the duration of the laser excitation and the choice of the metal transducer. As presented here, the use of 35 fs laser pulses and Ta results to a ~10 ps acoustic strain pulse (see Fig. 4), while laser pulses of several tens of picoseconds or longer would result to significantly longer strain pulses. The acoustic pulse duration could play a significant role in the control of the pressure field inside the crystal and the optimization of the γ-ray radiation.

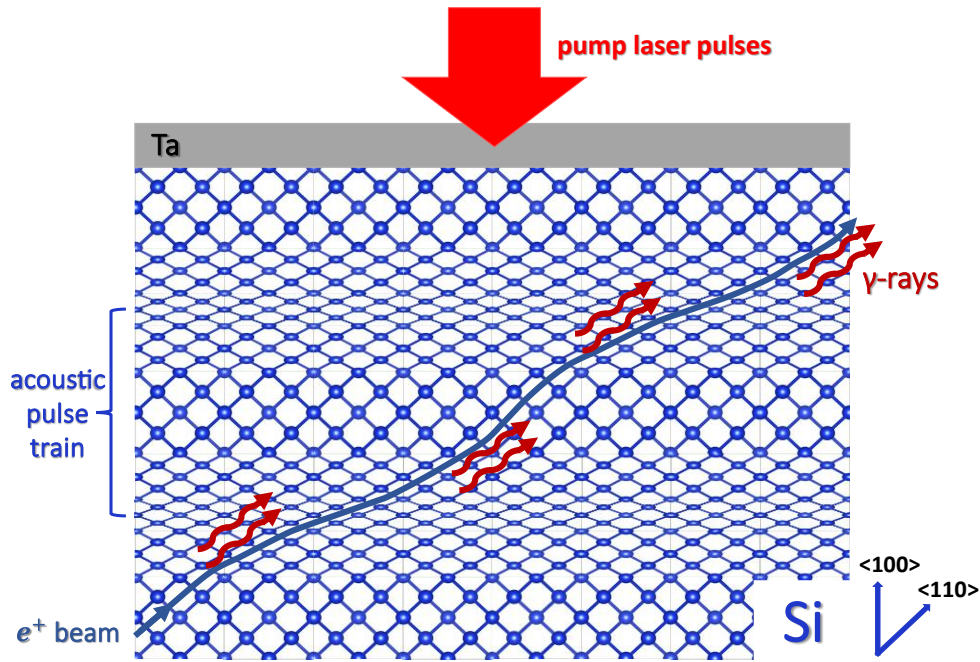

Fig. 6: Schematic representation of a laser-driven acoustic wave crystalline undulator based on Si <100>, and produced γ-ray radiation, where the $e^+$ beam propagates along the <110> direction.

## 5. Conclusions

In this work we have studied the photoacoustic properties of Ta thin films under ultrafast excitation using a combined experimental and computational method. Transition metals, and particularly Ti, have previously been shown to exhibit superior photoacoustic properties compared to noble metals, such as Ag. Here, a comparative evaluation of the photoacoustic transduction efficiency of Ta and Ti was presented using the pump-probe transient reflectivity method on 25 nm thick Ta and Ti coatings on Si substrates. The captured transient reflectivity signals were analyzed regarding the modulation depth of the Brillouin oscillations, which is directly related to the photoelastic change in the refractive index of Si due to the acoustic pulse, and hence, to the amplitude of the induced strain. In this respect, Ta was found to have a significantly higher transduction efficiency than Ti, a fact that was supported by computational evaluations of the electron and lattice temperatures and strains in the Ti/Si and Ta/Si samples, using FEA. The excellent photoacoustic properties of Ta, along with its favorable thermomechanical properties, such as the high yield strength and melting point, render it highly suitable for a wide variety of photoacoustic applications.

Moreover, the suitability of ultrafast photoacoustic transduction on thin metallic films for the generation of acoustic pulse trains with controlled spatiotemporal characteristics was presented. From the computational evaluations it was shown that, when there is an impedance mismatch between the metal and the substrate, the metallic film acts as an acoustic cavity in which acoustic energy oscillates between the metal's surface and the interface with the substrate. In each oscillation, part of the acoustic energy propagates into the substrate, shaping a train of strain pulses with a temporal periodicity equal to the roundtrip time in the acoustic cavity. It was discussed that by appropriate tuning of the metal thickness, the material impedances, and the characteristics of the optical radiation, the spatiotemporal characteristics of the pulses can be controlled.

Finally, a novel application of photoacoustically-generated pulse trains in crystalline materials was proposed for the development of CUs. Such devices enable the generation of tunable γ-rays via undulation of $e^-/e^+$ ultra-relativistic beams inside the channels formed by the lattice planes of the crystal. Here, it is proposed that the particular Ta/Si sample and excitation could be suitable for the SASP undulation regime; however, by tuning the photoacoustic transduction parameters, other undulation regimes can also be achieved.

## Declarations

### Funding

The authors KKa, MKK, EK, VD, MT, MB, NP acknowledge the financial support by the TECHNO-CLS project of the Horizon Pathfinder programme of the European Innovation Council (G.A. 101046458 — TECHNO-CLS — HORIZON-EIC-2021-PATHFINDEROPEN-01). The authors KKa, EK and KKo acknowledge funding by the Hellenic Foundation for Research and Innovation (H.F.R.I.) under the "2nd Call for H.F.R.I. Research Projects to support Post-Doctoral Researchers" (Project Number: 1336).

### Acknowledgement

This work was supported by computational time granted by the Greek Research & Technology Network (GRNET) in the National HPC facility ARIS-under project ID pr013024-LaMPIOS II.

### Conflict of interests

All authors declare that they do not have any conflict of interest related to this work.

### Author contributions

KKa, MKK, KKo, MB and NP set up the experiments and carried out the measurements, VB and EG prepared the samples. EK, VD, MT and NP developed the theoretical model. KKa prepared the first draft. All authors revised the article.